\documentclass[conference]{IEEEtran}
\IEEEoverridecommandlockouts
\usepackage{cite}
\usepackage{amsmath,amssymb,amsfonts}
\usepackage{algorithmic}
\usepackage{graphicx}
\usepackage{textcomp}
\usepackage{xcolor}
\def\BibTeX{{\rm B\kern-.05em{\sc i\kern-.025em b}\kern-.08em
    T\kern-.1667em\lower.7ex\hbox{E}\kern-.125emX}}

\definecolor{blue}{HTML}{1F77B4}
\definecolor{green}{HTML}{2CA02C}
\definecolor{orange}{HTML}{FF7F0E}
\definecolor{red}{HTML}{D62728}
    
\usepackage{url}
\usepackage[hidelinks]{hyperref}
\usepackage[caption=false,font=footnotesize]{subfig}
    
\begin{document}


\title{Blind Deep-Learning-Based Image Watermarking Robust Against Geometric Transformations
{\footnotesize
\thanks{This work was funded in part by the Research Foundation -- Flanders (FWO), IDLab (Ghent University -- imec), Flanders Innovation \& Entrepreneurship (VLAIO), the Flemish Government, and the European Union.
The computational resources (STEVIN Supercomputer Infrastructure) and services used in this work were kindly provided by Ghent University, the Flemish Supercomputer Center (VSC), the Hercules Foundation and the Flemish Government department EWI.}
}}

\author{\IEEEauthorblockN{Hannes Mareen, Lucas Antchougov, Glenn Van Wallendael, and Peter Lambert}
\IEEEauthorblockA{
Ghent University \--- imec, IDLab, Department of Electronics and Information Systems, Ghent, Belgium, \\
firstname.lastname@ugent.be, \url{https://media.idlab.ugent.be}
}
}

\maketitle

\begin{abstract}
Digital watermarking enables protection against copyright infringement of images.
Although existing methods embed watermarks imperceptibly and demonstrate robustness against attacks, they typically lack resilience against geometric transformations.
Therefore, this paper proposes a new watermarking method that is robust against geometric attacks.
The proposed method is based on the existing HiDDeN architecture that uses deep learning for watermark encoding and decoding. We add new noise layers to this architecture, namely for a differentiable JPEG estimation, rotation, rescaling, translation, shearing and mirroring.
We demonstrate that our method outperforms the state of the art when it comes to geometric robustness.
In conclusion, the proposed method can be used to protect images when viewed on consumers' devices.
\end{abstract}


\section{Introduction}
Image watermarking has applications such as copyright protection, authentication, and fingerprinting or traitor tracing~\cite{video-watermarking-overview, mareen2018novel}. It is typically important for the watermark to be robust. For example, the watermark should survive attacks such as compression and geometric transformations. Additionally, to not hinder the consumers' viewing experience, it is desirable for the watermark to be imperceptible. Moreover, for the detection process to be applicable on a broad scale (such as on consumer devices), it is also important for the watermark to be blind, meaning the watermarking can be detected without requiring the original image nor additional information.

HiDDeN (Hiding Data with Deep Networks) is an advanced, deep-learning-based image watermarking technique~\cite{hidden}. This technique involves the integration of multiple networks into one end-to-end Convolutional Neural Network (CNN). A first network aims to embeds a watermark imperceptibly into the image, and second network aims to accurately decode the embedded watermark. The decoding does not require the original content, making it a blind watermarking technique. Although HiDDeN demonstrated good performance, it is not robust against geometric transformations.

This paper proposes a novel watermarking method robust against geometric attacks, by extending the HiDDeN architecture~\cite{hidden} with noise layers that simulate geometric attacks.

\section{Proposed Method}
\label{sec:proposed}

\begin{figure}[!t]
  \centering
  \includegraphics[width=1.0\linewidth, trim=100 150 100 150]{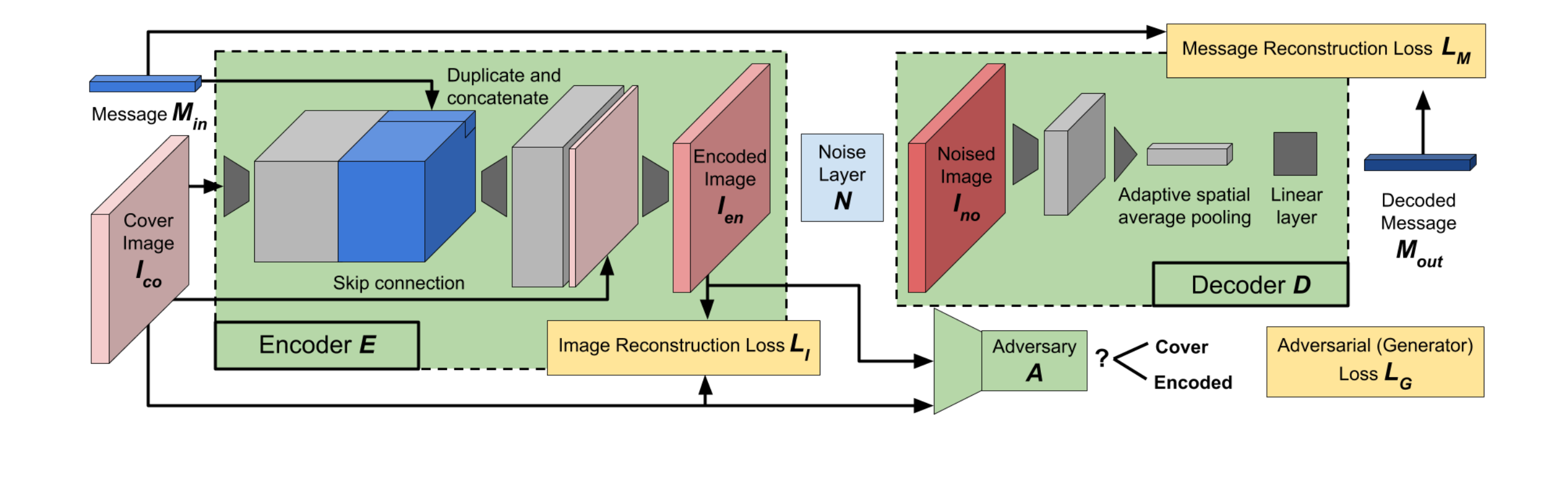}
  \caption{Architecture of HiDDeN~\cite{hidden}, consisting of an encoder and decoder, separated by noise layers. We include noise layers that simulate geometric transformations to provide robustness against geometric attacks.}
  \label{fig:method-hidden}
  \vspace{-0.3cm}
\end{figure}

We propose to extend the HiDDeN architecture~\cite{hidden} in order to make it robust against gemoetric distortions.
The HiDDeN architecture is visualized in Fig.~\ref{fig:method-hidden}, and consists of three networks to achieve this task: an encoder, decoder, and adversary.
The encoder network takes an image and a bit message as input and produces an output image with the message embedded into it.
The decoder network then takes the encoded image as input and returns a decoded message. The decoder is blind, i.e., it does not take any additional information as input.
The adversary network plays a crucial role by being trained to differentiate between watermarked and non-watermarked images, essentially acting as a discriminator.
This dynamic enables the encoder and decoder networks to learn how to ”outsmart” the adversary network, resulting in a substantial improvement in imperceptibility for the embedded watermark.
Using the adversary network results in a substantial improvement in imperceptibility~\cite{hidden}.

Between the encoder and the decoder networks, multiple so-called noise layers can be inserted. These layers simulate attacks on the watermark.
The originally proposed HiDDeN model~\cite{hidden} used the following noise layers: Crop, Gaussian Blur, Cropout, Dropout, and JPEG compression through JPEG-Mask \& JPEG-Drop.
We opted to not use Cropout and Dropout, as these use information from the unwatermarked image and are therefore not realistic attacks. Additionally, we replaced the JPEG-Mask and JPEG-Drop layers because we could not achieve good robustness against JPEG compression with them.
%
Instead,
we propose to use a JPEG noise layer on a differentiable approximation of JPEG compression called JPEG\textsubscript{diff}~\cite{jpeg-diff-paper} (instead of the JPEG-Mask and JPEG-Crop noise layers that the original method used).
JPEG\textsubscript{diff} works in a similar way as JPEG, with the difference that the rounding operation after quantization uses a rounding approximation instead. That is because the rounding operation $\lfloor I\rceil$ has a derivative of 0 nearly everywhere. The differentiable approximate rounding is done in the following way:
$ \lfloor I\rceil_{approx} = \lfloor I\rceil + (I - \lfloor I\rceil)^3$. This network is trained separately to simulate JPEG and is then used as a noise layer.

\begin{figure}[!t]
  \centering
  \includegraphics[width=1.0\linewidth,trim=70 60 70 50, clip=True]{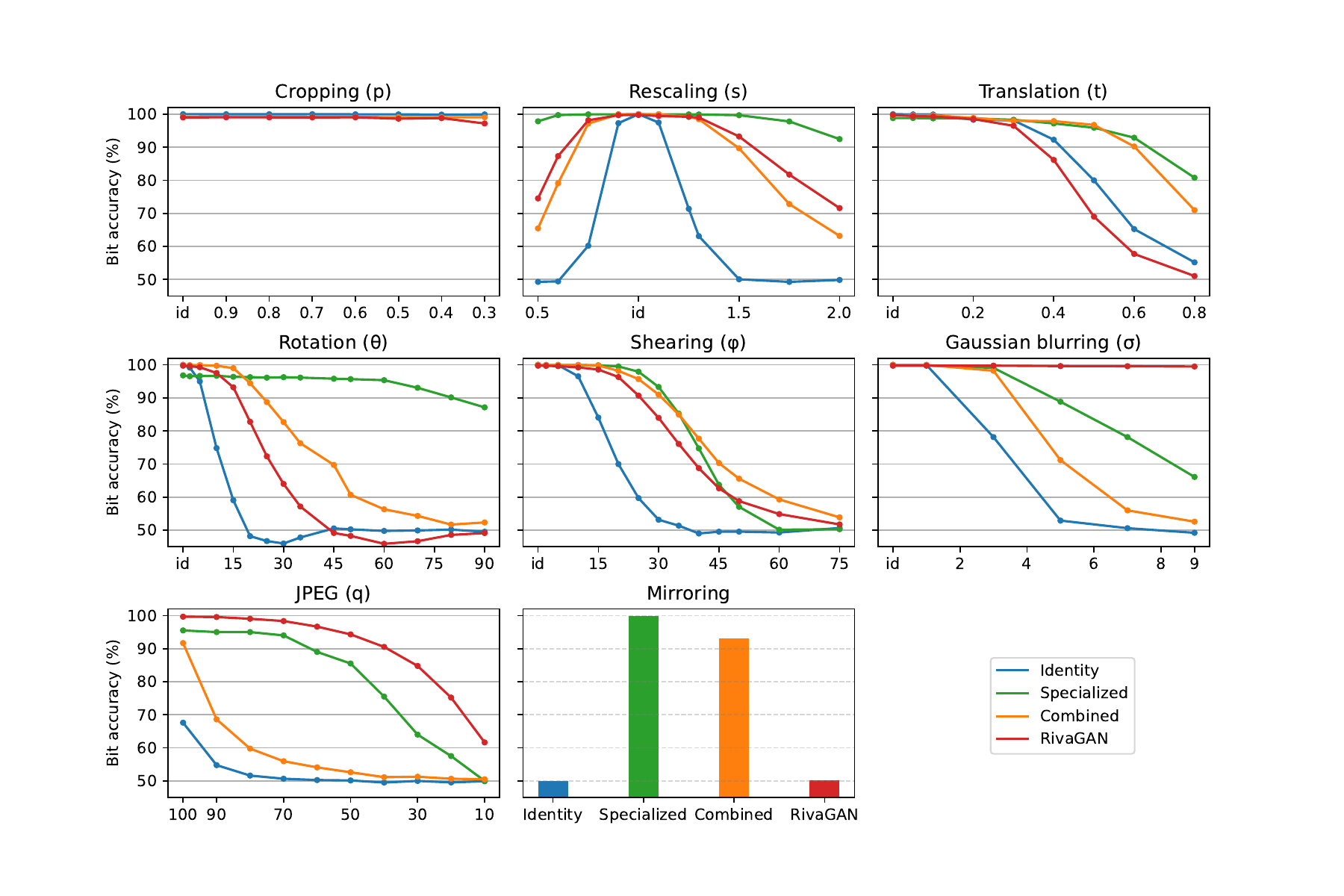}
  \caption{Robustness results.}
  \vspace{-0.3cm}
  \label{fig:evaluation-robustness}
\end{figure}

To provide robustness against geometric transformations, we add the following noise layers: Rescale (factor between 50\% and 200\%), Translate (5\% to 50\% of the image width and height), Rotate (angles between +/- 10 and 60 degrees), Shear (angles between +-/ 10 and 45 degrees), and Mirror. These operations are all differentiable. Additionally, we deleted the Crop noise layer, as we found that the base model (without noise layers) already provided sufficient robustness against cropping (see Section~\ref{sec:evaluation} and Fig.~\ref{fig:evaluation-robustness}).

\begin{table}
  \centering
  \caption{Imperceptibility results}
  \label{tab:evaluation-imperceptibility}
  \begin{tabular}{|c|c|c|c|c|}
    \hline
    \textcolor{blue}{\textbf{Identity}} & \textcolor{green}{\textbf{Rescale}} & \textcolor{green}{\textbf{Translate}} & \textcolor{green}{\textbf{Rotate}} & \textcolor{green}{\textbf{Shear}} \\
    \hline
    0.975 & 0.960 & 0.977 & 0.963 & 0.966 \\
    \hline
    \textcolor{green}{\textbf{Mirror}} & \textcolor{green}{\textbf{Gaussian Blur}} & \textcolor{green}{\textbf{JPEG\textsubscript{diff}}} & \textcolor{orange}{\textbf{Combined}} &  \textcolor{red}{\textbf{RivaGAN}} \\
    \hline
    0.980 & 0.963 & 0.963 & 0.930 & 0.977 \\
    \hline
  \end{tabular}
\end{table}

\section{Evaluation}
\label{sec:evaluation}
\subsection{Experimental Setup}
To create the training, validation, and testing, we selected 10000, 1000, and 1000 images from the COCO dataset~\cite{lin2014microsoft}, respectively. During training, random crops of $128\times 128$ pixels are taken, whereas testing was done on rescaled images of $512\times 512$ pixels (to evaluate imperceptibility) and $256 \times 256$ pixels (to evaluate robustness).
Additionally, we embedded randomly-generated 30-bit messages.
We train and evaluate an Identity model (no noise layers), 7 specialized models (single noise layer), and a Combined model (6 noise layers, selected at random during training). The Shear noise layer was not added in the Combined model, because the Rotate layer already sufficient provided robustness against shearing.
We additionally evaluate and compare against the state-of-the-art RivaGAN method~\cite{zhang2019robust}.

\subsection{Imperceptibility}
To measure the imperceptibility, we calculate the structured similarity index measure (SSIM) between the original image and its watermarked version. A higher value signifies a less perceptible watermark. Table~\ref{tab:evaluation-imperceptibility} gives average SSIM values for all methods. Although the combined models scores the lowest SSIM score (0.930), it 
remains an imperceptible result. Furthermore, RivaGAN exhibits an SSIM score of 0.977, performing similarly to most specialized models. We manually inspected the watermarked images and confirm that it is very hard (if not impossible) to spot artefacts at native resolution.
Example images are available on our website: \url{https://media.idlab.ugent.be/watermarking-blind-icce}.

\subsection{Robustness}
To measure the robustness, we evaluate the bit accuracy of the decoded messages for a range of attacks.
It can be observed that the identity model (blue line) performs relatively poorly against most attacks. The surprising exception is cropping, for which it maintains a bit accuracy close to 100\% for crops with ratio $p \geq 0.2$.
The specialized models (green lines) never fall below 65\% bit accuracy for the considered attacks.
The combined model shows significantly improved robustness compared to the identity model. Unsurprisingly, the combined model has a lower resilience than the specialized models for their targeted attacks, though. This is especially the case for JPEG compression and Gaussian Blurring.
Future work should investigate how to provide additional robustness in these cases.

In contrast, RivaGAN shows very good performance against JPEG compression and Gaussian blurring.
This demonstrates RivaGAN's ability to extract low-level features of images that can survive certain attacks without an explicit need for noise layers~\cite{zhang2019robust}.
However, most notably, our specialized and combined methods outperform RivaGAN when exposed to geometric attacks.

\section{Conclusion}
\label{sec:conclusion}
We presented a blind deep-learning-based watermarking method based on the HiDDeN architecture~\cite{hidden}, which uses noise layers to provide robustness against specific attacks.
We demonstrate that the embedded watermarks are imperceptible and, most notably, we outperform state-of-the-art methods in robustness against geometric attacks.
As such, the watermarking method can be used to protect copyright of images viewed on consumer electronic devices.


\bibliographystyle{IEEEtran}
\bibliography{IEEEabrv, main}

\end{document}